\documentclass[aps,prl,twocolumn,groupedaddress]{revtex4-1}
\usepackage{graphicx}
\usepackage{amsmath}
\usepackage{amssymb}
\usepackage{amsfonts}
\usepackage{subeqnarray}
\usepackage[utf8x]{inputenc}
\usepackage{epstopdf}
\usepackage{color}

\graphicspath{{Figures/},{images/}}

\begin{document}
\title{Deformation of a free interface pierced by a tilted cylinder}

%\author{Christophe Raufaste}
%\email[corresponding author : ]{Christophe.Raufaste@unice.fr}
%\affiliation{Laboratoire de Physique de la Matière Condensée, CNRS UMR 6622, Université de Nice Sophia-Antipolis, 06108 Nice, France}
%\author{Geoffroy Kirstetter}
%\affiliation{Laboratoire de Physique de la Matière Condensée, CNRS UMR 6622, Université de Nice Sophia-Antipolis, 06108 Nice, France}
%\author{Franck Celestini}
%\affiliation{Laboratoire de Physique de la Matière Condensée, CNRS UMR 6622, Université de Nice Sophia-Antipolis, 06108 Nice, France}
%\author{Simon Cox}
%\affiliation{Institute of Mathematics and Physics, Aberystwyth University, SY23 3BZ, UK}

\author{Christophe Raufaste}
\email[corresponding author : ]{Christophe.Raufaste@unice.fr}
\affiliation{Laboratoire de Physique de la Mati\`ere Condens\'ee, CNRS UMR 7336, Universit\'e de Nice Sophia-Antipolis, 06108 Nice, France}
\author{Geoffroy Kirstetter}
\affiliation{Laboratoire de Physique de la Mati\`ere Condens\'ee, CNRS UMR 7336, Universit\'e de Nice Sophia-Antipolis, 06108 Nice, France}
\author{Franck Celestini}
\affiliation{Laboratoire de Physique de la Mati\`ere Condens\'ee, CNRS UMR 7336, Universit\'e de Nice Sophia-Antipolis, 06108 Nice, France}
\author{Simon Cox}
\affiliation{Institute of Mathematics and Physics, Aberystwyth University, Ceredigion SY23 3BZ, UK}

%\pacs{47.55.-t}{Multiphase and stratified flows}
%\pacs{47.15.-x}{Laminar flows}
%\pacs{68.03.-g}{Gas-liquid and vacuum-liquid interfaces}

\date{\today}

\begin{abstract}
We investigate the interaction between an infinite cylinder and a free fluid-fluid interface governed only by its surface tension. We study the deformation of an initially flat interface when it is deformed by the presence of a cylindrical object, tilted at an arbitrary angle, that the interface ``totally wets''. Our simulations predict all significant quantities such as the interface shape, the position of the contact line, and the force exerted by the interface on the cylinder. These results are compared with an experimental study of the penetration of a soap film by a cylindrical liquid jet.
This dynamic situation exhibits all the characteristics of a totally wetting interface.
We show that whatever the inclination, the force is always  perpendicular to the plane of the interface, and its amplitude diverges as the  inclination angle increases. Such results should bring new insights in both fluid and solid mechanics, from animal locomotion to surface micro-processing.
\end{abstract}

%\pacs{47.55.-t, 47.15.-x, 68.03.-g}
%47.15.-x: laminar flows
%47.55.-t: Multiphase and stratified flows
%68.03.-g: Gas-liquid and vacuum-liquid interfaces
%\keywords{}

\maketitle

\section{Introduction}

Free fluid-fluid interfaces  produce fascinating patterns in nature. From the simple geometry of drops, bubbles and micelles \cite{DeGennes2003}, to the complex but regular arrangement of films inside a liquid foam \cite{Cantat2010},  the concept of interfacial energy allows us to rationalize the change of bulk properties at the boundaries between two immiscible fluids. In the case of liquid-gas interfaces, deformation of the interface by a solid object and the  subsequent measurement of the force exterted by one on the other  is still a widely-used technique to quantify the surface tension (for instance the Du No\"uy Ring and Wilhelmy plate methods \cite{Adamson1976}).
In many cases the object that pierces the interface is cylindrical,  and the deformation of a flat liquid-gas interface due to the contact of a solid rod is a situation of interest in both solid and fluid mechanics. 

The propulsion of insects that walk on water is possible because of the force between the interface and the rigid cylindrical legs \cite{Hu2003,Hu2005,Vella2008,Su2010}. If the latter are flexible, then a new class of so-called ``elasto-capillary'' or ``wet-hair'' problems \cite{Bico2004}  arise with major applications in nano- or microdevices. Thus  hairy surfaces, such as carbon nanotubes and biological filaments, are examples of the rich morphologies that arise from individual hair deformation and collapse \cite{Neukirch2007,Park2008, Blow2010,Tawfick2011} and of the patterns seen in assemblies of hairs \cite{Cohen2003,Chakrapani2004, Chiodi2010, Duan2010} in the presence of a meniscus. 
Related problems are encountered in surface processing, for example in fiber coating and in describing the wetting and cleaning properties of textiles \cite{Lorenceau2004,Mullins2007,Huang2009}; or more specifically in probing the wetting properties of  nanoscale droplets by deforming them with a cylindrical AFM tip \cite{Connell2002,Mullins2007}. 
Finally, the mutual force between a soap film and a cylindrical liquid object, namely a micro-jet, was also found to be strong enough to cause the jet to deviate and to control its trajectory \cite{KirstetterPreprint}.

In general, the rise of a meniscus is fast \cite{Clanet2002} compared to other relevant time-scales  (solvent evaporation, leg motion). Consequently, the equilibrium shape of the gas-liquid interface and the minimum-energy principle are usually used to account for observations.

Here, we use  Surface Evolver \cite{brakke92} simulations to study the deformation of an initially flat interface when it is pierced with an infinite cylinder. We report the effect of the inclination angle on the meniscus shape, the contact line position and on the force generated by the interaction. In the following we consider the case of a ``total wetting'' condition, meaning that the contact angle between the interface and the cylinder is zero. The results are general as long as the interface is governed only by  surface tension, meaning that they can be directly applied to interfaces between two immiscible liquids, boundaries between a liquid and a gas, and to most  2D dimensional lamellae, including soap films, Langmuir monolayers,  lipid bilayers and  biological membranes. The main control parameter is the inclination of the cylinder with respect to the plane of the unperturbed interface. The numerical study is compared with experiments in which a cylinder of liquid is brought into contact with a soap film; 
above a certain liquid velocity the results are in good agreement with the model, offering further insight into the dynamical interaction between a cylinder and a film.
As explained below, the more the cylinder is inclined, the greater the deformation of the meniscus and the greater the force exerted by the meniscus on the cylinder. This force is always perpendicular to the interface and its amplitude increases rapidly as the inclination  is increased, diverging as the angle approaches 90$^\circ$.

The manuscript is organized as follows: the numerical and experimental systems are first presented, then we describe and characterize the position of the contact line in the simulations, compare with the experiments, and show how to calculate the force on the cylinder from the position of the  contact line. We determine the force as a function of the inclination from the simulation results, and discuss the success of different models in predicting this dependence.

\section{Materials and Methods\label{Sec:MM}}

\begin{figure}%[!h]
a) Simulations
\begin{center}
\includegraphics[width=6cm]{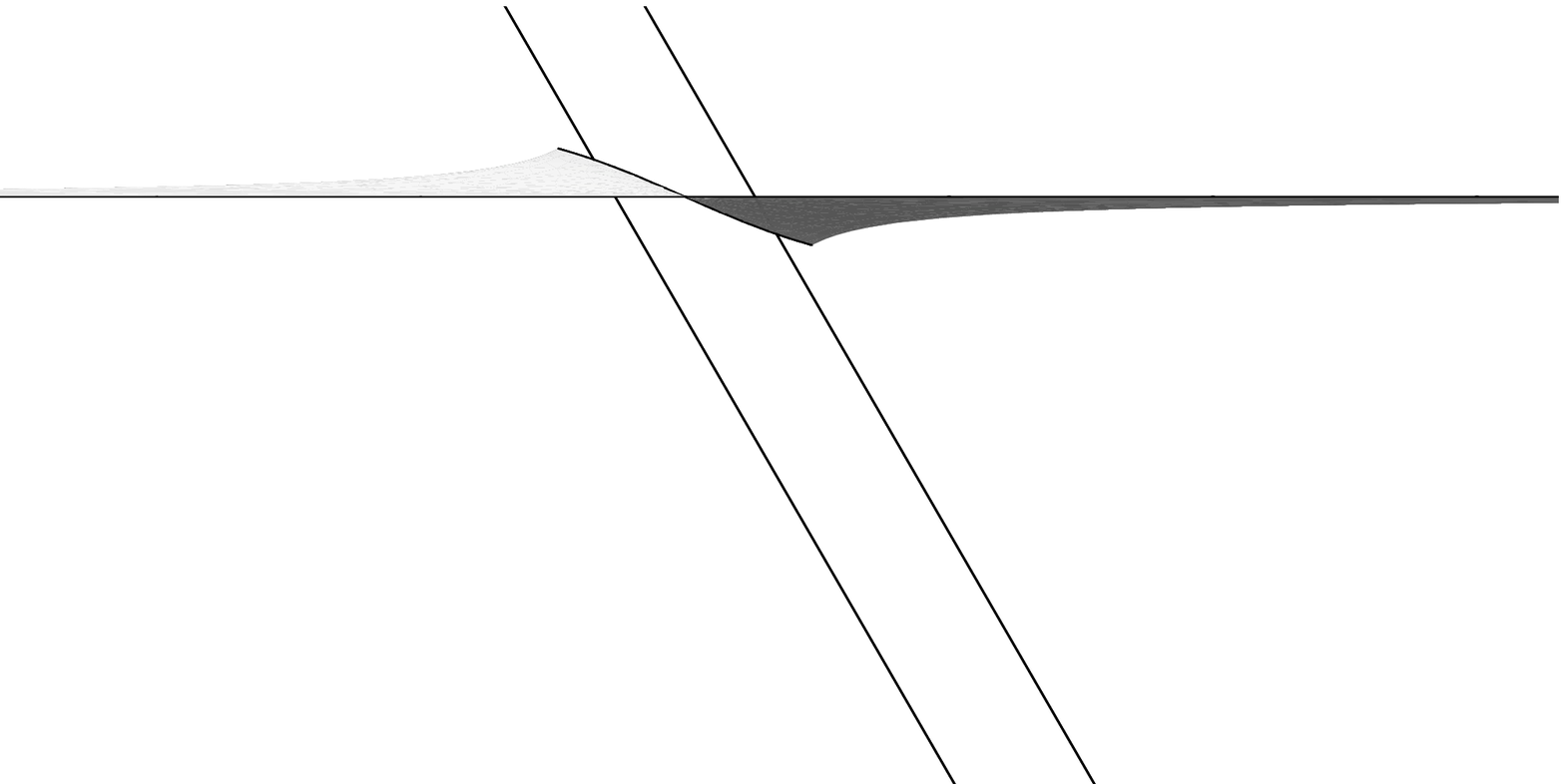}
\includegraphics[width=6cm]{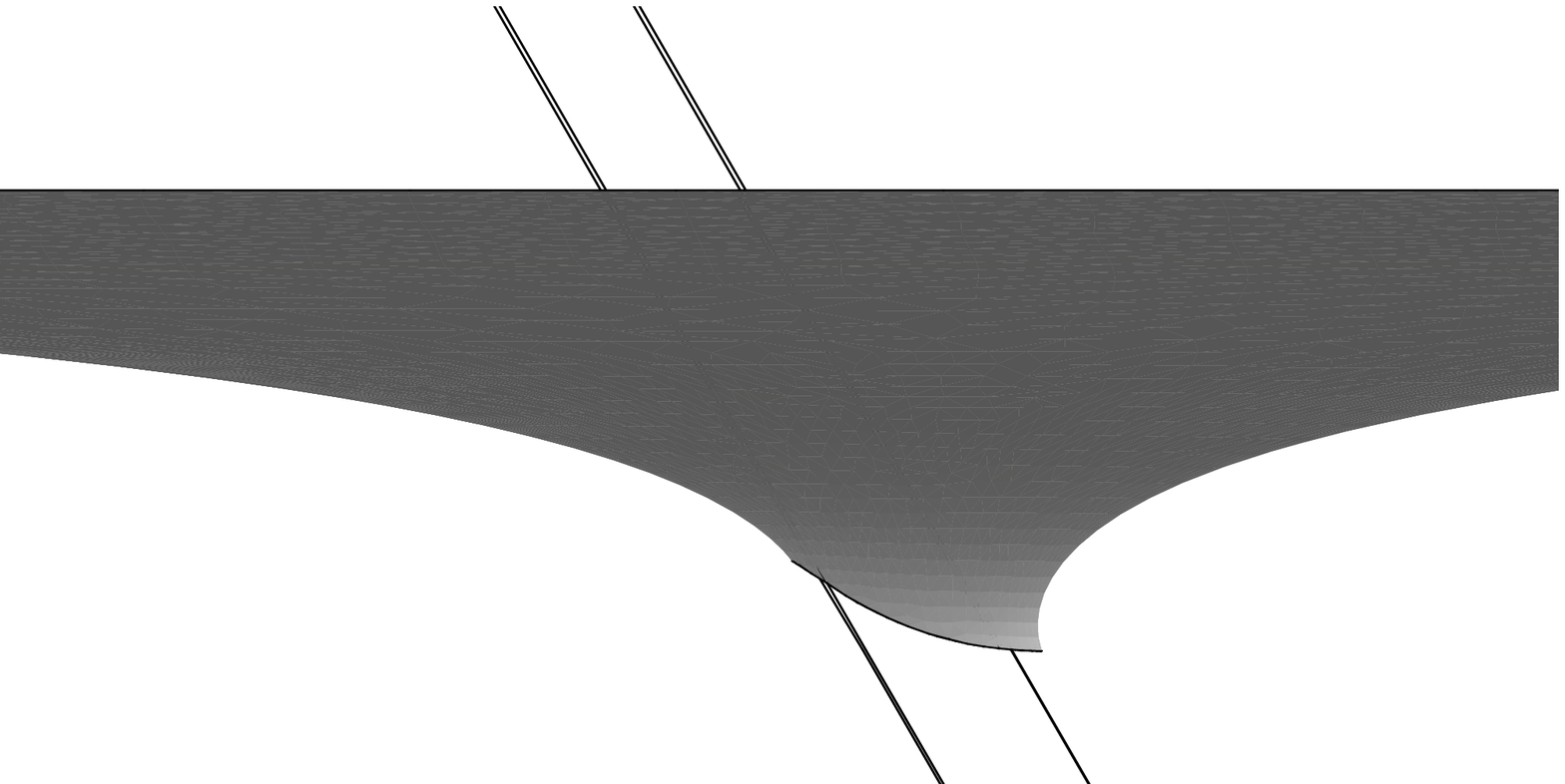}
\end{center}
b) Experiments
\begin{center}
\includegraphics[width=8.8cm]{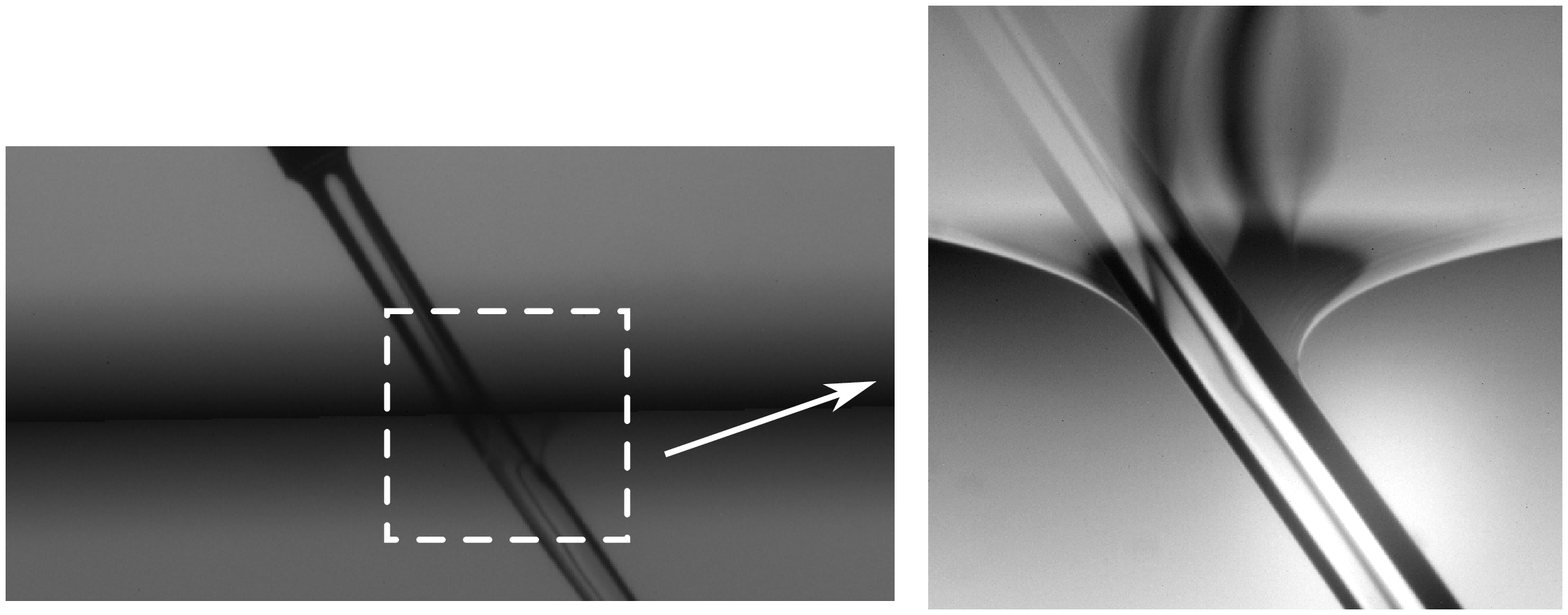}
\end{center}
\caption{\label{Fig:ImagesSimulExpe} Side view showing how the interface deforms. a) Simulation, with $\theta_i=30^\circ$, 
and an imposed contact angle of $90^\circ$ (top) or $0^\circ$ (bottom).
 b) Experiment, with $\theta_i=30^\circ$, $R=200\mu$m and $V=2 $m.s$^{-1}$.
}
\end{figure}

\subsection{Simulations}

We use the Surface Evolver \cite{brakke92} to calculate the shape of a single area-minimizing surface that is pierced by a rod at an inclination of $\theta_i$ to the vertical. All units are dimensionless, without loss of generality. We first create a planar surface in the $x-y$ plane with surface tension 1.005 in a fixed square frame of side-length 5.0 and a cylindrical rod, with axis in the vertical ($z$) direction and radius $R=0.1$, through its centre. To ensure that the rod is wetted, we include a virtual liquid film on the lower half of the rod with unit tension. The small difference in tension between the surface and this virtual film ensures that the contact angle is slightly greater than zero, alleviating numerical problems.

The surface is meshed with many small triangular facets. We tested two levels of refinement, the first with about 12,000 triangles, and the second with an extra 2000 triangles in the region where the rod pierces the surface; the refinement also controls the number of points around the contact line, giving 46 and 81 points respectively. Both triangulations give equivalent results (data not shown), which validates the robustness of the approach, and we show the more refined results below.

One iteration consists of tilting the rod by $3^\circ$ in the $x-z$ plane and finding the new minimum area configuration of the surface (Fig. \ref{Fig:ImagesSimulExpe}a). At each iteration, the location of the points along the contact line is recorded which, as we shall show below, is sufficient to calculate all quantities of interest.

\subsection{Experiments}

To counteract the effects of bouyancy (which dominate capillary forces at these length-scales), we consider a double interface formed by a soap film, with surface tension $2 \gamma$, rather than the single interface separating a liquid from air. Our conclusions remain valid for such an interface, but with half the value of the surface tension.

The force exerted on a static glass rod by a meniscus, with a contact angle equal to $90^\circ$, is zero. To generate a contact angle of $0^\circ$, we therefore use a laminar liquid jet directed through a soap film of the same composition, as described in \cite{KirstetterPreprint,Celestini2010}. The flow in the jet plays a subtle role: the liquid flow within the soap film is slow, so the film adopts a quasistatic equilibrium shape, yet the boundary conditions are governed by the dynamics of the jet. As we shall demonstrate below, friction causes the contact line to move downstream, and in the limiting case imposes the required contact angle on the soap film.

The liquid is a soap solution obtained by adding 5\% of commercial dish-washing liquid (Dreft, Procter \& Gamble) to deionised water.  The surface tension of the soap solution is $\gamma = 26.2 \pm 0.2$ mN.m$^{-1}$ and its density is $\rho = 10^{3}$kg.m$^{-3}$. As reported elsewhere \cite{KirstetterPreprint}, this situation is stable and  the meniscus forms an equilibrium shape which is recorded by a digital camera (Fig. \ref{Fig:ImagesSimulExpe}b). The incident jet is characterized by its incident angle $\theta_i$, velocity $V$ and radius $R$. 

The Weber number, $We = \rho V^2 R/\gamma$ is used to estimate the relative contributions of inertia and capillary forces. Within our experimental parameter range, the Reynolds number based on the jet characteristics is always significantly larger than unity.

\section{Results\label{Sec:Results}}

\subsection{Shift of the contact line}

\begin{figure}%[!h]
\includegraphics[width=8.5cm]{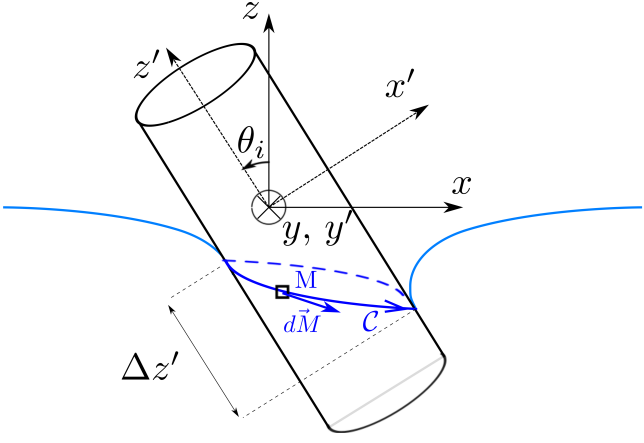}
\caption{\label{Fig:JetFilmFrame} Schematic representation of the system.
}
\end{figure}

\begin{figure}[!h]
\setlength{\unitlength}{1cm} 
\centering
\begin{picture}(8.5,8)(0.0,0.0)
\put(-0.5,0){\resizebox{9.5cm}{!}{\includegraphics[width=9.5cm]{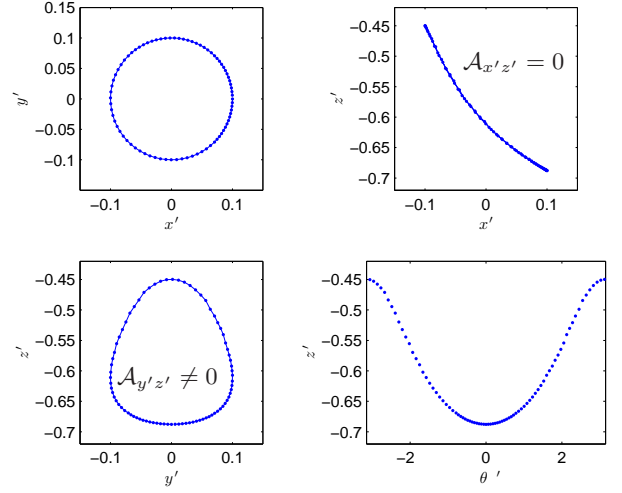}}}
\put(6.2,5.8){\textbf{$\mathcal{A}_{x^\prime z^\prime}=0$}}
\put(1.6,1.6){\textbf{$\mathcal{A}_{y^\prime z^\prime}\neq0$}}
\end{picture}
\caption{\label{Fig:TripleLinePosition} The contact line projected on different planes for a simulation with for $\theta_i = 30^\circ$, in the reference frame of the cylinder $\mathcal{R}^\prime$.
}
\end{figure}

The shape and position of the contact line where the interface meets the cylinder is sketched in Fig. \ref{Fig:JetFilmFrame}. The reference frame of the initially flat interface is $\mathcal{R}$ (O$xyz$), with O$x$ horizontal, and  the reference frame of the cylinder is $\mathcal{R}^{\prime}$ (O$x^\prime y^\prime z^\prime$); they are related by
\begin{subeqnarray}
\label{Eq:xzxzprime}
\slabel{Eq:Ann2EquationQuasistatique_1}
x^\prime & = & \cos(\theta_i) x  + \sin(\theta_i) z, \\
y^\prime & = & y,\\
z^\prime & = & -\sin(\theta_i) x + \cos(\theta_i) z.
\end{subeqnarray}
All calculations are performed in the reference frame of the cylinder, $\mathcal{R}^{\prime}$, using cylindrical coordinates ($r^\prime$, $\theta^\prime$, $z^\prime$), so that any point M has
$\vec{OM} = x^\prime \vec{e}_{x^\prime} + y^\prime \vec{e}_{y^\prime} + z^\prime \vec{e}_{z^\prime} 
= r^\prime \vec{e}_{r^\prime} +\theta^\prime \vec{e}_{\theta^\prime} + z^\prime \vec{e}_{z^\prime}$, with
\begin{subeqnarray}
x^\prime & = & r^\prime \cos(\theta^\prime)\\
y^\prime & = & r^\prime \sin(\theta^\prime)\\
\vec{e}_{r^\prime} &= & \cos(\theta^\prime) \vec{e}_{x^\prime} + \sin(\theta^\prime) \vec{e}_{y^\prime}\\
\vec{e}_{\theta^\prime} &= & -\sin(\theta^\prime) \vec{e}_{x^\prime} + \cos(\theta^\prime) \vec{e}_{y^\prime}.
\end{subeqnarray}
By symmetry,  the  contact line (Fig. \ref{Fig:TripleLinePosition}) is given by
\begin{subeqnarray}
r^\prime & = & R\\
z^\prime & = & z^\prime(\theta^\prime),
\end{subeqnarray}
with $\theta^\prime \in [-\pi,\pi]$.

\begin{figure}[!h]
\centering
\includegraphics[width =10cm]{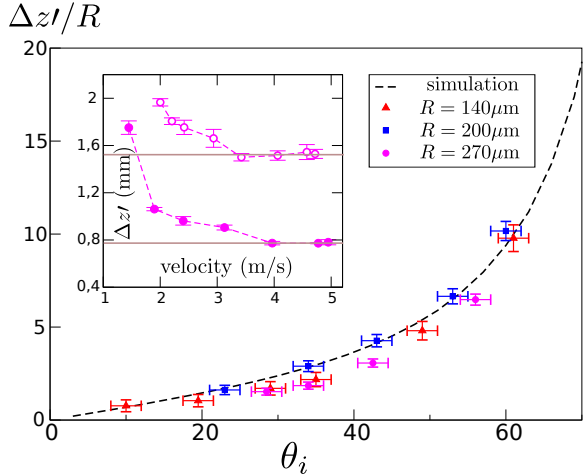}
\caption{\label{Fig:ResExp}  The quantity $\Delta z^\prime/R$ {\it versus} $\theta_i$ in simulations (dashed line) and in experiments (different radii $R = 140$, 200 and 270 $\mu$m are shown as circles, triangles and squares respectively). Insert:  experimental data for $\Delta z^\prime$ {\it versus} the incident velocity $V$ for a radius of 270 $\mu$m and two incident angles: 39$^{\circ}$(filled circles) and 49$^{\circ}$(open circles). The two solid lines indicate the values to which the curves saturate. }
\end{figure}

The position of the contact line in the simulation with an incident angle of $\theta_i = 30^{\circ}$, referred to $\mathcal{R}^{\prime}$, is shown in Fig. \ref{Fig:TripleLinePosition}. The {\em deviation} of the contact line is defined by $\Delta z^\prime = z^\prime(\theta^\prime = \pi) - z^\prime(\theta^\prime = 0)$. $\mathcal{A}_{x^\prime y^\prime}$, $\mathcal{A}_{x^\prime z^\prime}$ and $\mathcal{A}_{y^\prime z^\prime}$ denote the absolute values of the  area enclosed by the contact line when projected on the (O$x^\prime y^\prime$),  (O$x^\prime z^\prime$) and (O$y^\prime z^\prime$) planes respectively. By symmetry, $\mathcal{A}_{x^\prime y^\prime} = \pi R^2$ and $\mathcal{A}_{x^\prime z^\prime} = 0$. $\mathcal{A}_{y^\prime z^\prime}$ and $\Delta z^\prime$  are equal to zero when $\theta_i = 0$ and both increase monotonically as $\theta_i$ increases. Fig. \ref{Fig:ResExp} shows how $\Delta z^\prime/R$ depends on $\theta_i$ in the simulation (dashed line).

Side views of the meniscus were used to investigate experimentally the effect of $R$, $\theta_i$ and $V$ on the meniscus shape and perform measurements of  $\Delta z^\prime$.  For a given set of parameters ($R$, $\theta_i$), $\Delta z^\prime$ was found to decrease as $V$ increases (insert to Fig. \ref{Fig:ResExp}) until saturating above a limiting velocity. This marks the onset of  the dynamical ``total wetting'' condition (see Discussion). In this regime, the meniscus shape is found to be independent of the jet velocity and the following measurements were performed at sufficiently high velocity to ensure that $\Delta z^\prime$ reached its saturation value.

Fig. \ref{Fig:ResExp} shows $\Delta z^\prime/R$ as a function of $\theta_i$, illustrating the agreement between simulations and experiments. This scaling allows the collapse of all  experimental measurements (several $R$ and $\theta_i$ values) on the curve found from the simulation. 

\subsection{Force measurement}

Assuming total wetting, it is possible to infer the force on the cylinder from the contact line position.

\begin{figure}[!h]
\centering
\includegraphics[width=8.5cm]{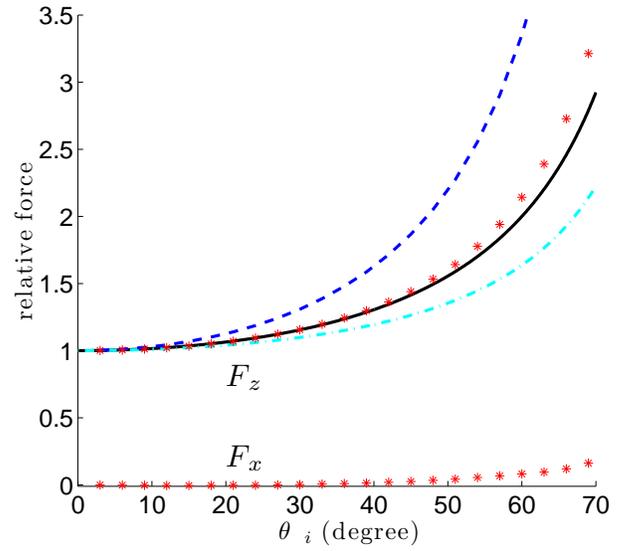}
\caption{\label{Fig:ForceAngle} The different force components $F_x$ and $F_z$ are normalised by $2\gamma \pi R$. Simulations are shown as red stars. Three models are tested for $F_z$. Black solid line: vertical force model, $F_z = 2\gamma \pi R /\cos(\theta_i)$. Blue dashed line: total perimeter model. Cyan dashed-dotted line: flat interface model.
}
\end{figure}

We denote by M($R$, $\theta^\prime$, $z$($\theta^\prime$)) a point on the contact line and $d\vec{M}$ an infinitesimal variation of position along the contact line (Fig. \ref{Fig:JetFilmFrame}): $d\vec{M} = Rd\theta^\prime \vec{e}_{\theta^\prime} + dz^\prime \vec{e}_{z^\prime} = d\theta^\prime \left( R \vec{e}_{\theta^\prime} + \frac{dz^\prime}{d\theta^\prime} \vec{e}_{z^\prime} \right)$.
Knowing that the local force is perpendicular to $d\vec{M}$ and to $\vec{e}_{r^\prime}$ (total wetting condition), the local force  $d\vec{F}$ applied to $d\vec{M}$ is
\begin{equation}
\label{Eq:LocalForce_dF}
 d\vec{F} = \gamma \vec{e}_{r^\prime} \wedge d\vec{M}.
\end{equation}
The prefactor $\gamma$ holds for single interfaces and should be replaced by $2\gamma$ without loss of generality for double (e.g. air-liquid-air) interfaces such as soap films. The total force that the interface exerts on the cylinder  is
\begin{equation}\label{Eq:ForceIntegralExpression}
\vec{F} = \gamma \oint_\mathcal{C} \vec{e}_{r^\prime} \wedge d\vec{M},
\end{equation}
where $\mathcal{C}$ represents the oriented contour around the contact line.
After some manipulation, described in the Appendix, an exact expression for the force  in the reference frame $\mathcal{R}$ of the interface is obtained:
\begin{subeqnarray}
\label{Eq:ForceR}
\slabel{Eq:Fxzxzprime2}
F_x & = & 2\gamma \pi R \left( \cos(\theta_i) \frac{\mathcal{A}_{y^\prime z^\prime}}{2\pi R^2}  - \sin(\theta_i) \right), \\
F_y & = & 0, \\
F_z & = & 2\gamma \pi R \left( \sin(\theta_i) \frac{\mathcal{A}_{y^\prime z^\prime}}{2\pi R^2} + \cos(\theta_i)  \right),
\end{subeqnarray} 
where $\mathcal{A}_{y^\prime z^\prime}$ is the absolute value of the projected area of the contour in the (O$y^\prime$ $z^\prime)$  plane.
Force measurements are then obtained numerically using the contour profiles calculated in the simulations and the results are shown in Fig. \ref{Fig:ForceAngle}. For the whole range of $\theta_i$ tested,  $F_x$ is small compared to $F_z$, suggesting that the force acts vertically (in the reference frame of the interface).

The vertical component of force can be compared to three different models (Fig. \ref{Fig:ForceAngle}):
\begin{itemize}
\item vertical force model: from the observation above, we might assume $F_x=0$. Consequently Eq. \ref{Eq:Fxzxzprime2} gives $\frac{\mathcal{A}_{y^\prime z^\prime}}{2\pi R^2}  = \tan(\theta_i)$  and thus 
\begin{equation}
F_z = 2\gamma \pi R /\cos(\theta_i).
\label{eq:mainresult}
\end{equation}

\item total perimeter model: the length of the contact line, $\mathcal{P}$, which is measured in the simulations, might determine the vertical force: $F_z = \gamma \mathcal{P}$.

\item flat interface model: $F_z$ could depend on the perimeter of the ellipse that results from projecting the cross-section of the cylinder on to the unperturbed interface ((O$x y$) plane): $F_z = \gamma \mathcal{P}_{ellipse}$. This ellipse has semi-axes of length $R$ and $R/\cos(\theta_i)$. An approximation to the perimeter of this ellipse (which is valid to better than 5\% in our parameter range) leads to $F_z = 2\gamma \pi R \left(1+ \displaystyle \frac{2}{\pi}\left(\displaystyle \frac{1}{\cos(\theta_i)} -1  \right) \right)$. 
\end{itemize}

\section{Discussion\label{Sec:Discussion}}

\subsection{Comparison of models}
The deformation  of an interface by an object of characteristic size $L$ should result in a force that scales as $\gamma L$. A rod has two well-separated characteristic lengths (radius and axial length) so the force could, in principle, range over several orders of magnitude. In the case of an infinite rod, one of the characteristic lengths is infinite, which explains why the force diverges as the rod becomes horizontal.

It is clear that the vertical force model gives the best fit, and the good agreement with the measurements extends up to very large inclination angles.
The total perimeter model is plotted as it represents an upper bound for $F_z$, since it assumes that all local forces $d\vec{F}$ (Eq. \ref{Eq:LocalForce_dF}) are vertical in the reference frame of the interface. The flat interface model underestimates the force measurements. {\it A priori} we might have expected the opposite, as for the total perimeter model (which also accounts for vertical local forces), but it significantly underestimates the real perimeter of the contact line: the presence of the cylinder perturbs the contour and there is a significant shift in the vertical direction $\Delta z = z( \theta^\prime = \pi) - z( \theta^\prime = 0)$ which increases its perimeter. The vertical force model works well, and it would be interesting to see if this is also the case for ``partially wetted'' contact. 

Based on an energy balance  \cite{Neukirch2007} a vertical force model  has also been used in the study of an elastic plate that pierces an interface, enabling the authors to infer a $1/\cos(\theta_i)$ dependence to explain the different buckling morphologies of the plate under the effect of capillary forces.
Expression (\ref{eq:mainresult}) could also be helpful in understanding the detachment of liquid droplets from fibres \cite{Lorenceau2004,Mullins2007,Huang2009}, in which a drop is pierced by a rod non-axisymmetrically. This complicated geometry has two contact lines that prevent the drop from being detached by gravity, and the rod is locally inclined to the surface of the drop. No exact expressions for the interface shape or contact line position are available, so that in \cite{Lorenceau2004} an expression for the force was based on experimental data, with no divergence as the inclination angle increased. This is in contrast to our $1/\cos(\theta_i)$ dependence for an infinite planar interface, which is explained by noting that our expression would fail as soon as the deformation of the contact lines became comparable to the distance between them or to the drop size. In that case, the total capillary force might saturate, which supports the maximal capillary force expression frequently used to describe detachment above a given drop size.

\section{Conclusion}

The wetted contact of a free interface with a cylindrical object was studied both numerically and experimentally. Quasistatic simulations give the deformation of the interface and the position of the contact line, in good agreement with experiments on a soap film in contact with a cylinder of liquid. 
Above a limiting velocity, the liquid jet applies a dynamical ``total wetting'' condition to the interface, with what we might call a dynamically-generated capillary force,  which is also of paramount importance in explaining the refraction of the jet, as described in Kirstetter et al. \cite{KirstetterPreprint}.
From the contact line position, the force exerted by the interface on the object (and {\em vice-versa}) is calculated. We observe that whatever the inclination of the cylinder, the force remains vertical and its amplitude is given by $F_z = 2\gamma \pi R /\cos(\theta_i)$. This expression should be useful in numerous problems in both fluid and solid mechanics, 
and in future work it could be extended to study the partially wetted case, validated by the sort of simulations described here.

\section{Appendix: Expression for the force}
\label{App:ForceMeasurements}

Expanding the integral for the force  (Eq. \ref{Eq:ForceIntegralExpression}) leads to
\begin{eqnarray*}
\vec{F} /\gamma & = & \oint_\mathcal{C} \vec{e}_{r^\prime} \wedge \left(R \vec{e}_{\theta^\prime}+ \frac{dz^\prime}{d\theta^\prime} \vec{e}_{z^\prime}\right) d\theta^\prime \\
& = & \oint_\mathcal{C} \left(R \vec{e}_{z^\prime}- \frac{dz^\prime}{d\theta^\prime} \vec{e}_{\theta^\prime}\right) d\theta^\prime \\
& = & \oint_\mathcal{C} \left(R \vec{e}_{z^\prime}+ \sin(\theta^\prime) \frac{dz^\prime}{d\theta^\prime} \vec{e}_{x^\prime} - \cos(\theta^\prime) \frac{dz^\prime}{d\theta^\prime} \vec{e}_{y^\prime}\right) d\theta^\prime,
\end{eqnarray*}
which gives the force components along the three axes of the  reference frame of the cylinder $\mathcal{R}^{\prime}$ (O$x^\prime y^\prime z^\prime$):
\begin{eqnarray*}
F_{x^\prime}/\gamma & = & \oint_\mathcal{C} \left( \sin(\theta^\prime) \frac{dz^\prime}{d\theta^\prime} \right) d\theta^\prime\\
F_{y^\prime}/\gamma & = & \oint_\mathcal{C} \left(- \cos(\theta^\prime) \frac{dz^\prime}{d\theta^\prime} \right) d\theta^\prime\\
F_{z^\prime}/\gamma & = & \oint_\mathcal{C} R  d\theta^\prime.
\end{eqnarray*}
After some calculation, the contour integrals can be written in cylindrical or cartesian coordinates as:
\begin{eqnarray*}
F_{x^\prime}/\gamma & = & - \oint_\mathcal{C} \left( z^\prime \cos(\theta^\prime)  \right) d\theta^\prime \\
			&=& - \oint_\mathcal{C}  z^\prime d(\sin(\theta^\prime))\\ 
			&=& - \displaystyle\frac{1}{R} \oint_\mathcal{C}  z^\prime dy^\prime   \\
F_{y^\prime}/\gamma & = & - \oint_\mathcal{C} \left( z^\prime \sin(\theta^\prime)  \right) d\theta^\prime\\
			&=&  \oint_\mathcal{C}  z^\prime d(\cos(\theta^\prime))\\ 
			&=&  \displaystyle\frac{1}{R} \oint_\mathcal{C}  z^\prime dx^\prime\\
F_{z^\prime}/\gamma & = &   R \oint_\mathcal{C}  d\theta^\prime \\
			& = &  2\pi R.
\end{eqnarray*}
The last expression for each component shows that $F_{x^\prime}$ and $F_{y^\prime}$ are directly related to the projected areas of the contact line in the (O$y^\prime$ $z^\prime)$  and (O$x^\prime$ $z^\prime$) planes  respectively. With $\mathcal{A}_{y^\prime z^\prime}$ and $\mathcal{A}_{x^\prime z^\prime}$ the absolute values of these oriented areas, as defined previously, we have by symmetry (Fig. \ref{Fig:TripleLinePosition}) that $\mathcal{A}_{x^\prime z^\prime} = 0$ and $\oint_\mathcal{C}  z^\prime dy^\prime < 0$. This leads to the final expression for the force in the reference frame of the cylinder $\mathcal{R}^{\prime}$ (O$x^\prime y^\prime z^\prime$):
\begin{eqnarray*}
F_{x^\prime} & = & 2\gamma \pi R \frac{\mathcal{A}_{y^\prime z^\prime}}{2\pi R^2} \\
F_{y^\prime} & = & 0\\
F_{z^\prime} & = & 2\gamma \pi R
\end{eqnarray*}
Writing the force components in the  reference frame of the interface $\mathcal{R}$ can be done using the change-of-coordinates  formulae (Eq. \ref{Eq:xzxzprime}), which then gives Eq. \ref{Eq:ForceR}.

\end{document}